\documentclass[conference]{IEEEtran}
\IEEEoverridecommandlockouts
\usepackage{amsmath,amssymb,amsfonts}
\usepackage{algorithmic}
\usepackage{graphicx}
\usepackage{textcomp}
\usepackage{xcolor}
\usepackage{tabularx, multirow}
\usepackage{array}
\usepackage{url}
\newcolumntype{S}{>{\hsize=0.5\hsize\linewidth=\hsize}X}
\newcolumntype{L}{>{\hsize=1.5\hsize\linewidth=\hsize}X}

\usepackage[style=ieee,doi=false,isbn=false,url=false,eprint=false,maxnames=4]{biblatex}
\begin{document}

\title{{\normalsize Author preprint, to appear in the 2026 North American Power Symposium (NAPS), accepted April 2026.}\\
Quantifying the Resilience Benefits of Undergrounding a Circuit with Utility Data\\
\thanks{
Funding from the Iowa State University Electric Power Research Center and USA NSF grants 2153163 and 2429602 is gratefully acknowledged.}
}

\author{\IEEEauthorblockN{Arslan Ahmad, Ian Dobson, and Anne Kimber}
\IEEEauthorblockA{
Iowa State University\\
Email: arslan@iastate.edu, dobson@iastate.edu, akimber@iastate.edu}
}

\maketitle

\begin{abstract}
We leverage historical outage data to quantify the resilience benefits of undergrounding a circuit. The historical performance of the overhead circuit is compared to the performance if the circuit had been undergrounded in the past. The number of outages, customers affected, outage duration, and customer hours lost are used as metrics to quantify the benefits of undergrounding. Results show 75\% and 78\% reductions in customer hours lost per year for two selected circuits, as well as a significant reduction in the average number of outages and customers affected per year, highlighting the advantages of undergrounding. The benefits of investments that result in 10\% faster outage restoration are also calculated by rerunning history with the faster restoration included.
\end{abstract}


\section{Introduction}

Undergrounding a circuit significantly reduces the number of outages. It can increase restoration times, but is overall very beneficial to resilience, including resilience to extreme winds. 
Therefore, resilience considerations should be taken into account when a utility proposes undergrounding a circuit. 
This work aims to quantify these resilience benefits using available utility data to help make the case for undergrounding.
The new method compares historical outage records from when the circuit was overhead with the outages that would have occurred if the circuit had been underground in the past. 
This method has the advantages of being driven by real data and being likely to persuade stakeholders. Appealing to the benefits that would have been gained for specific past extreme events that stakeholders remember is likely more persuasive than predicting future benefits for rare extreme events at some unspecified time in the future.

This paper makes the following contributions:
\begin{itemize}
    \item Introduces a rerunning history method to quantify the resilience benefits of undergrounding distribution circuits using real utility outage data rather than forward-looking simulations.
    \item Quantifies undergrounding benefits using event-based resilience metrics 
    that are directly interpretable by utilities.
    \item Demonstrates the method on two real distribution circuits, showing large reductions in customer-dependent impact metrics.
    \item Demonstrates the method's generalizability by extending it to quantify the resilience benefits of non-infrastructural investments, specifically faster restoration.
\end{itemize}

\section{Literature Review}

Undergrounding parts of the electrical power distribution systems has been a topic of interest for both utilities and consumers in the United States. Various surveys and reports have shed light on the attitudes, preferences, and challenges associated with this transition. While aesthetics and improved reliability are often cited as benefits, the high cost of undergrounding is a significant barrier to its widespread adoption.

A 2012 survey by the Edison Electric Institute (EEI) of 1,003 residential customers found that 39\% already had underground service \cite{hallEEI13}. This suggests a significant portion of the population is already familiar with the benefits.
Furthermore, another similar survey by EEI indicated that a significant number of customers were willing to bear additional costs on their electricity bills to have electric wires in their neighborhoods underground. Specifically, 34\% of respondents were willing to pay between 1\% and 10\% more, while 26\% were willing to pay even higher premiums \cite{hallEEI13}.

While public perception seems favorable, cost remains a significant hurdle. Reports from entities like Navigant \cite{navigant2005LIPA} and studies focused on specific regions like Florida \cite{brown2007Infrasource, brown2007Infrasource2, le2008Quanta} have provided insights into the practical implications and feasibility of undergrounding projects. These reports highlight the significantly higher initial construction costs of undergrounding compared to overhead lines.
The average cost per mile for new underground electric construction can range from \$25,000 to \$1.5 million. Whereas the average cost per mile for overhead construction can range from \$15000 per mile (for single-phase taps in rural communities) to \$250000 per mile for three-phase lines \cite{virginia2005Commission}.
It is essential to note that relocating 100\% of the existing overhead utility distribution lines and placing all new utility distribution lines underground may not be feasible. Additionally, converting existing overhead lines to underground is generally more expensive than building new underground circuits from scratch \cite{virginia2005Commission}.
However, partially underground circuits do not achieve the full resilience benefits of undergrounding, as the remaining overhead sections still have lower resilience.

Understanding the pros and cons of both overhead and underground circuits is crucial in evaluating the feasibility and desirability of undergrounding projects. The trade-offs between overhead and underground systems broadly fall under three heads: initial construction costs, reliability, and operation \& maintenance (O\&M).  
The major advantage of overhead distribution circuits, in most cases, is significantly lower initial construction costs. Additionally, individual faults on overhead circuits can be repaired more quickly. Some experts believe that overhead circuits have a longer lifespan, with new underground circuits lasting at least 30 years and overhead circuits lasting 40 years or more. Overhead circuits also offer greater flexibility for circuit reconfiguration and can withstand overloads more readily.

On the other hand, the primary advantages of underground circuits are improved aesthetics and overall improved reliability and resilience. 
As per the 2021 Post‐Construction Report of the Wisconsin Public Service Commission (WPSC) System Modernization and Reliability Project (SMRP), a 95\% improvement in SAIDI (137-minute reduction compared to the SAIDI values from the 2008 through 2011 period when the corresponding circuits were overhead) was observed during 2021 on the underground circuits \cite{wpsc22}.
In its 2025 Strategic Underground Program Annual Report, the Virginia Electric and Power Company (VEPC) stated that it had converted 269 miles of overhead tap lines to underground by the end of 2024. 
As a result of this undergrounding, they saw a 99.6\% reduction in SAIDI (2.25 vs. 556 minutes), 99.2\% reduction in SAIFI (0.0085 vs. 1.0441), and 50.6\% reduction in CAIDI (263 vs. 532 minutes)\cite{vepc25}. These results include not only the number of affected customers directly connected to the underground line sections, but also the downstream customers who experienced an outage due to a fault in the underground section.

Florida Power \& Light (FPL) implemented an SSUP pilot program in which they started converting most of their overhead distribution laterals to underground. Owing to the performance of this pilot, it was expanded and continued to a permanent distribution lateral hardening Program. As a result of undergrounding, only 4\% of the total underground laterals (3767 out of 103384) faced a power outage during Hurricane Irma. Whereas, during the same hurricane, 24\% of the total overhead laterals (20341 out of 84574) faced power outages. FPL experienced, on average, a 73\% reduction in SAIDI across five years (2017 to 2021) for underground facilities compared to their overhead distribution facilities. Even the SAIDI for underground facilities was 54\% less than that of hybrid facilities (a mix of overhead and underground) \cite{fpl22}.
A 10\% increase in a system's underground line miles is correlated with a 14\% reduction in annual interruption durations across the U.S \cite{Larsen16}.

The average number of underground outages is 69\% less than overhead outages and the average duration of underground outages is 14\% longer than overhead outages \cite{hallEEI13}.
For non-storm events, the average repair time of underground outages is 57\% longer than overhead outages (2.8 hours vs. 4.4 hours) \cite{doc10}.
Although a specific fault on an underground circuit may take longer to locate and repair, underground systems fail less frequently, and the average customer outage time (averaged across all customers) is generally shorter for most underground systems.
Undergrounding also eliminates most momentary interruptions.

Furthermore, underground circuits have overall low operation and maintenance costs because undergrounding virtually eliminates the need for tree-trimming, eliminates vehicular accidents with utility poles, reduces some electrical hazards, and nearly eliminates the need for extensive restoration efforts after catastrophic storms \cite{lbnl24}.
A report by the Virginia Commission \cite{virginia2005Commission} emphasizes that cost (initial construction cost, annual O\&M costs, annual savings from a reduction in the costs due to less tree-trimming, post-storm restoration, vehicle accidents, and lost revenue from unmet demand) and reliability (post-storm restoration and lost revenues from unmet demand) are the two major deciding factors.

The decision to underground electrical power distribution systems requires careful consideration of various factors. While public perception seems receptive, the high initial cost remains a significant challenge. A comprehensive analysis that considers all the aspects of reduced maintenance, improved reliability, and reduced outages is crucial for informed decision-making.

Distribution system component failure rate models can be tuned based on real-world distribution system data \cite{zhang2007TDEI}.

\section{Overall Method}
\begin{figure}[ht]
    \centering
    \includegraphics[width=0.48\textwidth]{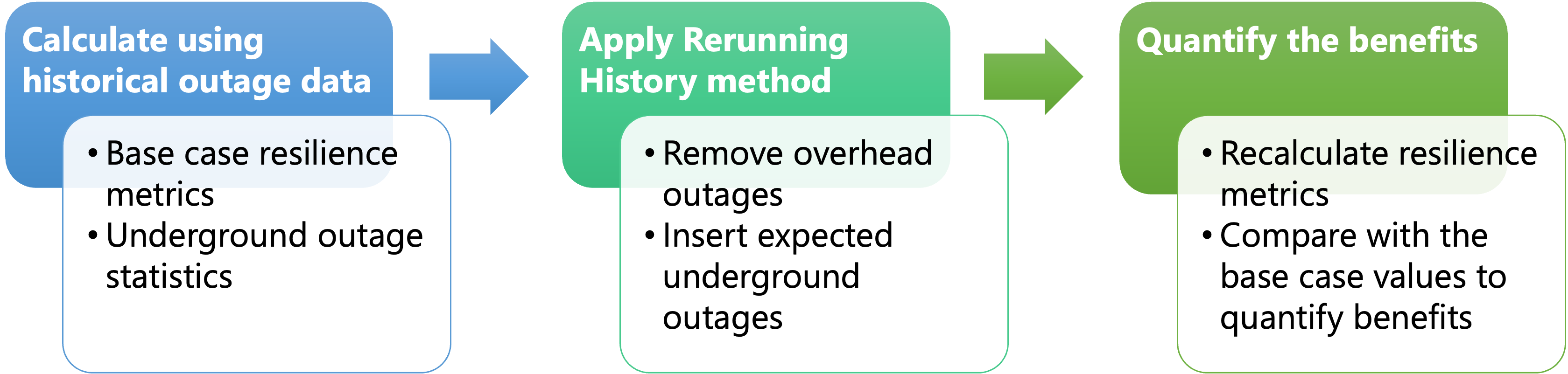}
    \caption{Method to quantify undergrounding benefits}
    \label{fig:flowDiagram}
\end{figure}

A specific overhead circuit is chosen and we aim to quantify the benefits if 80\% of the circuit had been undergrounded in the past\footnote{
80\% is only used to demonstrate the method; any value can be used as desired by the utility.}
The key steps of the method are shown in Fig.~\ref{fig:flowDiagram}.

Step 1 processes the historical outage data to identify the full range of events that occurred during the data period. 
Events can be small, involving only a few outages, or larger resilience events driven mainly by extreme weather. 
Then, the base-case resilience metrics for all events are calculated to quantify the circuit's performance when it was overhead. 
The resilience metrics include the number of outages, the number of customers affected, the duration of outages, and the total customer hours lost.
Outages that occurred on the already underground sections of the system are used to calculate typical statistics of underground outages, such as the average outage duration and average outage rate. These statistics are later used as input to the rerunning history method in step 2.

Step 2 considers the same circuit and applies the rerunning history method to estimate its performance if 80\% of the circuit had been undergrounded in the past.
80\% of the overhead outages are randomly removed from the circuit's data, and underground outages, characterized by the typical underground outage statistics calculated in step 1, are inserted.

Step 3 calculates the resilience metrics for the modified underground circuit and compares them with the base-case resilience metrics from step 1 to quantify the benefits of undergrounding.

A detailed explanation of the method is given in section \ref{section:UndergroundingDetails}.

Apart from the benefits of undergrounding, the benefits of investments leading to a 10\% faster restoration of outages are also calculated using the rerunning history technique to demonstrate another application of rerunning history. Its details are explained in section \ref{section:FastRestorationDetails}.

\section{Quantifying Undergrounding Benefits} \label{section:UndergroundingDetails}
The proposed method is explained in detail in the subsequent subsections, beginning with the assumptions made.

\subsection{Assumptions}  \label{section:assumptions}
\begin{itemize}
    \item We assume that both the primary and secondary distribution of the selected circuit is to be underground, i.e., all the way from the substation to the end consumer.
    \item We exclude the scheduled outages from the analysis and only consider the unscheduled or unplanned outages.
    \item We assume that 80\% of the circuit is to be undergrounded. However, the same process can be easily applied to different percentages or for analyzing small sections of a circuit or multiple circuits. Furthermore, undergrounding primarily eliminates exposure-related failures (such as wind and vegetation), and the random removal of outages is an approximation due to the unavailability of cause-level mapping. This method can easily be extended to cause-conditioned removal.
    \item Statistics of typical underground outages are calculated from historical underground outages observed in the underground sections of different circuits. Due to the unavailability of the exact dates when those sections were underground, they were assumed to have been underground since 2001.
\end{itemize}

The selected circuits (CIRCUIT1 and CIRCUIT2) do not have any underground sections. The length and details of outages (after excluding the scheduled outages) on the selected circuits are given in Table \ref{tab:selectedCircuitsDetails}.

\begin{table}[!ht]
    \caption{Details of outages on the selected circuits}
    \label{tab:selectedCircuitsDetails}
     \begin{tabularx}{\columnwidth}{X c c c c}
         Circuit&Length&Total&First&Last\\
         Code&(miles)&Outages&Outage&Outage\\
         \hline
         CIRCUIT1     & 10.15   & 46    & 06-14-2001   & 09-16-2020 \\
         CIRCUIT2     & 5.96    & 40    & 11-29-2001   & 04-22-2018 \\
         \hline
     \end{tabularx}
\end{table}
 
\subsection{Data Filtering}
We gratefully acknowledge Alliant Energy Corporation for generously providing data and advice for this analysis.
The data range spans almost 21.5 years: from January 2001 up to and including June 2022. This corresponds to the time of occurrence of the first and last outage in the dataset, i.e., January 6, 2001, and June 8, 2022, respectively.

The dataset contains a total of 804 outages. These include distribution, transmission, and substation outages. A breakdown of ``System Type" is given in table \ref{tab:sysTypeBreakup}. Due to the scope of the current analysis, only the 759 distribution system outages (both overhead and underground) are retained, and the rest are excluded. 
In addition to the ``System Type" information, ``Outage Cause" information is also available for each outage. It divides the data into different outage cause categories based on the causes recorded for each outage. We used the outage cause codes ``Sched Construction\_Intent", ``Sched Maintenance\_Intent", ``Sched Util Work\_Intent", and ``Govt Req/Pub\_Safety\_Intent" to identify the scheduled outages in the data. Since scheduled outages are pre-planned, they have different impacts and occurrence behaviors compared to unscheduled outages. Therefore, the 269 scheduled distribution system outages are also excluded. The analysis is performed on the remaining 490 outages. 

\begin{table}[!ht]
    \caption{Breakdown of outages data according to ``System Type"}
    \label{tab:sysTypeBreakup}
    \begin{tabularx}{\columnwidth}{X c c}
        System type     & Total outages     & Scheduled outages\\
        \hline
        Distribution - Overhead     & 735       & 259   \\
        Substation                  & 28        & 14   \\
        Distribution - Underground  & 24        & 10    \\
        Transmission - Overhead     & 11        & 5     \\
        Not Reported                & 4         & 0   \\
        Not an Outage               & 2         & 0     \\
        \hline
    \end{tabularx}
\end{table}

\subsection{Calculate Underground Outage Statistics} \label{section:outageStats}
Let $\bar{\lambda}_{ug}$ be the {\sl mean annual outage rate per mile} for underground outages, $\bar{\Delta t}_{ug}$ be the {\sl mean duration per outage} of underground outages, and $\bar{c}_{ug}$ be the {\sl mean number of customers affected per outage} in underground outages.
We estimate $\bar{\lambda}_{ug}$ and $\bar{\Delta t}_{ug}$ from the historical underground outage data of all the underground circuit segments in the distribution system.
There are 13 circuits with underground sections (Table~\ref{tab:ugCircuits}). 24 underground outages occurred on those circuits, 14 of which are unscheduled and used to estimate $\bar{\lambda}_{ug}$ and $\bar{\Delta t}_{ug}$ as:
\begin{align}
\bar{\lambda}_{ug} &= 0.12 ~\mbox{outages/mile/year}\label{lambdaug}\\
\bar{\Delta t}_{ug} &= 2.35~\mbox{hours}
\label{delatatug}
\end{align}
The estimates (\ref{lambdaug}) and (\ref{delatatug}) could be improved with more extensive underground circuit outage data or by using formulas for underground circuit reliability.

\begin{table}[!ht]
    \caption{Details of existing underground segments in circuits}
    \label{tab:ugCircuits}
    \begin{tabularx}{\columnwidth}{>{\raggedright\arraybackslash}X >{\centering\arraybackslash}c
    >{\centering\arraybackslash}X >{\centering\arraybackslash}X c c}
        &Total&Underground&&Total&Annual\\
        Circuit&Length&Length&Underground&UG&Outage\\
        Name&(miles)&(miles)&Outages&Years&Rate\\
        \hline
        CIRCUIT3   & 17.03 & 8.44  & 0 & 1.49   & 0.00   \\
        CIRCUIT4   & 11.50 & 8.79  & 0 & 1.49   & 0.00   \\
        CIRCUIT5   & 27.15 & 9.00  & 2 & 1.49   & 0.15   \\
        CIRCUIT6   & 47.50 & 29.50 & 2 & 3.58   & 0.02   \\
        CIRCUIT7   & 21.35 & 2.91  & 0 & 3.58   & 0.00   \\
        CIRCUIT8   & 17.74 & 2.09  & 0 & 3.58   & 0.00   \\
        CIRCUIT9    & 3.99  & 0.25  & 2 & 20.01  & 0.39   \\
        CIRCUIT10    & 6.95  & 1.18  & 2 & 20.01  & 0.08   \\
        CIRCUIT11    & 5.28  & 0.66  & 0 & 17.93  & 0.00   \\
        CIRCUIT12    & 4.67  & 0.09  & 0 & 17.93  & 0.00   \\
        CIRCUIT13    & 5.59  & 0.19  & 3 & 17.93  & 0.87   \\
        CIRCUIT14    & 17.13 & 1.09  & 0 & 17.93  & 0.00   \\
        CIRCUIT15    & 24.15 & 0.07  & 0 & 17.93  & 0.00   \\
        CIRCUIT16   & 14.35 & 0.07  & 3 & 17.93  & 0.12*   \\
        \hline\\[-0.9em]
        \multicolumn{5}{r}{$\bar{\lambda}_{ug}$(outages/mile/year):} &0.12\\[0.4em]
        \multicolumn{6}{l}
        {\footnotesize *Calculated value appears to be an outlier and is replaced with the average}
    \end{tabularx}
\end{table}

To estimate the {\sl mean number of customers affected per outage} in underground outages, we use the outage data of the circuits proposed for undergrounding (CIRCUIT1 and CIRCUIT2). 
This is because, even after undergrounding the circuit, the spatial distribution of customers on the circuit typically remains unchanged. Therefore, if a random outage occurs on the circuit after undergrounding, it will affect the same number of customers as an outage on the overhead circuit. 
And since these are the proposed circuits for undergrounding analysis, estimating the {\sl mean number of customers affected per outage} from these gives accurate results:
\begin{align}
\mbox{CIRCUIT1}\quad \mbox{mean }  \bar{c}_{ug} = 31~\mbox{customers}\\
\mbox{CIRCUIT2}\quad \mbox{mean } \bar{c}_{ug} = 41~\mbox{customers}
\end{align}
However, due to the highly skewed distribution of the number of customers, as shown in Fig~\ref{fig:ohCustomersPDF}, using mean values largely underestimates the undergrounding benefits in terms of customer-dependent metrics such as the number of customers affected and the customer hours lost. That is why we use the {\sl median number of customers affected per outage} instead of the {\sl mean number of customers affected per outage}:
\begin{align}
\mbox{CIRCUIT1}\quad \mbox{median }\tilde{c}_{ug} = 2~\mbox{customers}\\
\mbox{CIRCUIT2}\quad \mbox{median } \tilde{c}_{ug} = 7~\mbox{customers}
\label{cug}
\end{align}
It should be noted that the end results of the analysis are sensitive to the parameters $\bar{\lambda}_{ug}$, $\bar{\Delta t}_{ug}$, and $\tilde{c}_{ug}$, and therefore their accurate estimation is important to get accurate results.

\subsection{Calculate Overhead Resilience Metrics} \label{section:calculateMetrics}
We calculate resilience metrics using the historical outage data of the circuits proposed for undergrounding. 
We group outages into resilience events and calculate metrics for each event using the methods of \cite{CarringtonPESGM20}. 
This results in 46 and 38 events of different sizes for the CIRCUIT1 and CIRCUIT2 circuits, respectively. 
To quantify the average annual historical resilience performance of a circuit, we sum the values of each resilience metric for all events and divide the sum by the total number of years. 
These metrics serve as the base-case values and are compared with post-undergrounding metrics to quantify the benefits of undergrounding for each circuit.

We have used the following resilience metrics for each event in this analysis:
\begin{itemize}
    \item Number of outages
    \item Number of customers affected
    \item Customer hours
    \item Outage hours
\end{itemize}
In addition to using resilience metrics, reliability metrics such as SAIDI, SAIFI, CAIDI, and ASAI can also be used to quantify the benefits of undergrounding on the circuit's reliability using this method.

\subsection{Expected Underground Outages}
There are no underground outages on the two circuits selected for undergrounding.
Assuming that 80\% of the circuit is to be undergrounded, we remove all the overhead outages from this circuit except 20\% overhead outages chosen by random sampling (sampling without replacement). 
To ensure robustness, we repeat this process 2000 times and take the average of the results.
Using the $\bar{\lambda}_{ug}$ calculated in (\ref{lambdaug}), we calculate the number of outages per year that would have occurred in the selected circuit, using 20\% of the corresponding circuit lengths, $l$, given in Table~\ref{tab:selectedCircuitsDetails}, as:
\begin{align}
   n_{ug}=\bar{\lambda}_{ug}\times 0.2 \times l  
\end{align}

The annual total duration, number of customers affected, and customer hours of these outages are $n_{ug}\times\bar{\Delta t}_{ug}$, $n_{ug}\times\bar{c}_{ug}$, and $n_{ug}\times\bar{\Delta t}_{ug}\times\bar{c}_{ug}$, respectively.
We assume that underground outages do not bunch together; therefore, they will result in single outage events. 
This simplifies the calculation of the resilience metrics, and we obtain the annual values of the metrics as shown in Tables~\ref{tab:resultsCircuit1} and \ref{tab:resultsCircuit2}.

\section{Quantifying Faster Restoration Benefits} \label{section:FastRestorationDetails}
Similar to investments in undergrounding projects, we can apply the rerunning history method to quantify the resilience benefits of investments to improve the restoration process of outages. Some examples of such investments are:
\begin{itemize}
    \item Up-skilling existing crews and/or hiring more crews to complete the restoration process faster
    \item Equipping the crews with more and better resources
    \item Improving the post-event restoration planning with automation
    \item Improving the stockpiles of critical components
\end{itemize}
If such investments had been made, the restoration rates for outage events would have improved, resulting in earlier restoration completion. 
To demonstrate the effects of faster restoration investments on the resilience metrics, we assume investments that would have resulted in a 10\% faster restoration.
We rerun history by updating the restoration times of outages in the data using the method detailed in \cite{AhmadPS24}.
The results are shown in Table~\ref{tab:resultsFasterRestoration}.

\section{Results} \label{section:results}
The benefits of undergrounding 80\% of a distribution circuit are shown for two different circuits in Table \ref{tab:resultsCircuit1} and Table \ref{tab:resultsCircuit2}.
The resilience metrics are divided by the total number of years to yield the average annual benefit for easier interpretation. The percentage reduction in each metric is also shown.

\begin{table}[!ht]
    \caption{Benefits of undergrounding 80\% of CIRCUIT1 circuit}
    \label{tab:resultsCircuit1}
     \begin{tabularx}{\columnwidth}{X c c c}
        &&After 80\%&\\
        &Base Case&Undergrounding&\\
        Metric&(per year)&(per year)&Reduction\\
        \hline
        Number of Outages      & 2.3     & 1.4      & 39\%    \\
        Customers Affected     & 72.0    & 15.9     & 78\%   \\
        Customer Hours         & 184.2   & 40.2     & 78\%    \\
        Outage Hours           & 4.93    & 3.19     & 35\%     \\
        \hline
     \end{tabularx}
 \end{table}

\begin{table}[!ht]
    \caption{Benefits of undergrounding 80\% of CIRCUIT2 circuit}
    \label{tab:resultsCircuit2}
     \begin{tabularx}{\columnwidth}{X c c c}
        &&After 80\%&\\
        &Base Case&Undergrounding&\\
        Metric&(per year)&(per year)&Reduction\\
        \hline
        Number of Outages      & 2.2     & 0.94     & 58\%\\
        Customers Affected     & 91.4    & 20.0     & 78\%\\
        Customer Hours         & 118.9   & 30.3     & 75\%\\
        Outage Hours           & 4.99    & 2.18     & 56\%\\
        \hline
     \end{tabularx}
 \end{table}

\begin{table}[!ht]
    \caption{Benefits of investments leading to 10\% Faster Restorations}
    \label{tab:resultsFasterRestoration}
     \begin{tabularx}{\columnwidth}{ 
   >{\raggedright\arraybackslash}X 
   >{\centering\arraybackslash}X 
   >{\centering\arraybackslash}X }
         Metric&Base Case*&Reduction\\
         \hline
         Restore Duration       & 13.1 hours    & 9.8\%\\
         Event Duration         & 17.52 hours   & 4.0\%\\
         Customer Hours         & 600.56        & 3.4\%\\
         Outage Hours           & 61.11         & 4.5\%\\
         \hline
         \multicolumn{3}{l}
        {\footnotesize *Average values of all events with two or more outages}
     \end{tabularx}
 \end{table}

The results in Table~\ref{tab:resultsCircuit1} show that if the utility had changed 80\% of the CIRCUIT1 circuit from overhead to underground in January 2001, they would have seen a 39\% reduction in the number of outages per year on average on that circuit. The customer hours per year would have decreased by 78\% on average.
Similarly, the results in Table~\ref{tab:resultsCircuit2} show even higher reduction in all of the metrics if 80\% of the CIRCUIT2 circuit was underground.

Table~\ref{tab:resultsFasterRestoration} shows the improvements in different resilience metrics if investments had been made to decrease the restoration duration of outages by 10\%. Such investments would have resulted in a 3.4\% reduction in the customer hours of an event on average.

\section{Estimating the Typical Values of Outages}
\subsection{Duration of Outages}
\begin{figure}[ht]
    \centering
    \includegraphics[width=0.48\textwidth]{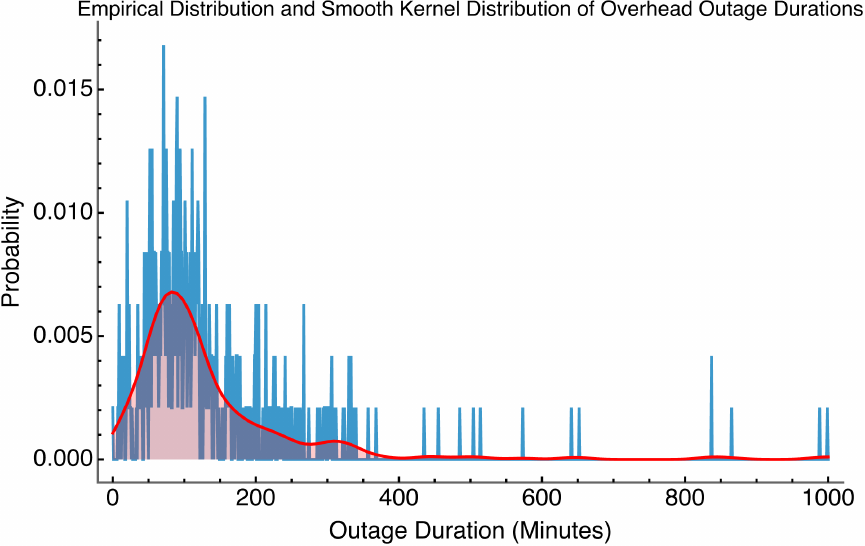}
    \caption{Empirical (blue) and estimated (red) probability density plot of the durations of unscheduled overhead outages}
    \label{fig:ohDurationPDF}
\end{figure}

Figure \ref{fig:ohDurationPDF} shows the empirical and estimated (estimated using a Gaussian Kernel) probability density of the durations of unscheduled overhead outages. Although the maximum outage duration in the data is 15752 minutes, the plot range is limited to 1000 minutes to highlight the major density area on the left side of the plot. The probability density is highly right-skewed, with a skewness coefficient of 8.4, and exhibits a long right tail. Fig. \ref{fig:ohDurationPDF} shows that almost 93\% of the probability lies below 400 minutes. In other words, there is a 93\% chance that an outage in the overhead distribution system will last less than 400 minutes. However, there is a 7\% chance that it will last more than 400 minutes!

The estimated probability density is unimodal, with a mode of 82 minutes. This means that a random outage in an overhead distribution system is most likely to last 82 minutes.
However, the arithmetic mean of the overhead outage durations is 335.8 minutes, which is far from this value and considerably overestimates the most common value.
This is a consequence of a highly right-skewed distribution.
Therefore, the arithmetic mean doesn't correctly represent the typical value of underground outage duration, and a more suitable statistic should be used instead.
One possible candidate is the median, which is 103 minutes in this case. 

\begin{figure}[ht]
    \centering
    \includegraphics[width=0.48\textwidth]{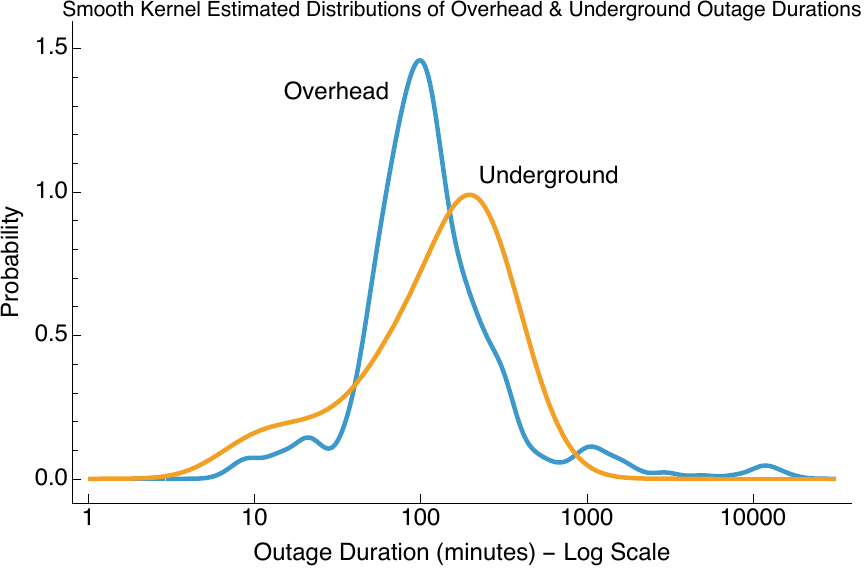}
    \caption{Estimated probability density of the log of the durations of unscheduled overhead and underground outages}
    \label{fig:ohAndUgDurationPDF}
\end{figure}

The unscheduled underground outage durations exhibit a similar probability behavior. However, there are only 14 unscheduled underground outages across all circuits in the data, and more data is needed to obtain accurate results. To compare overhead and underground outage durations, we transform the data using base-10 logarithms and estimate the probability density using a Gaussian kernel. 
The result is shown in Fig.~\ref{fig:ohAndUgDurationPDF}. We see that the typical duration of underground outages is higher than that of overhead outages.

\subsection{Number of Customers Affected in Outages}
\begin{figure}[ht]
    \centering
    \includegraphics[width=0.48\textwidth]{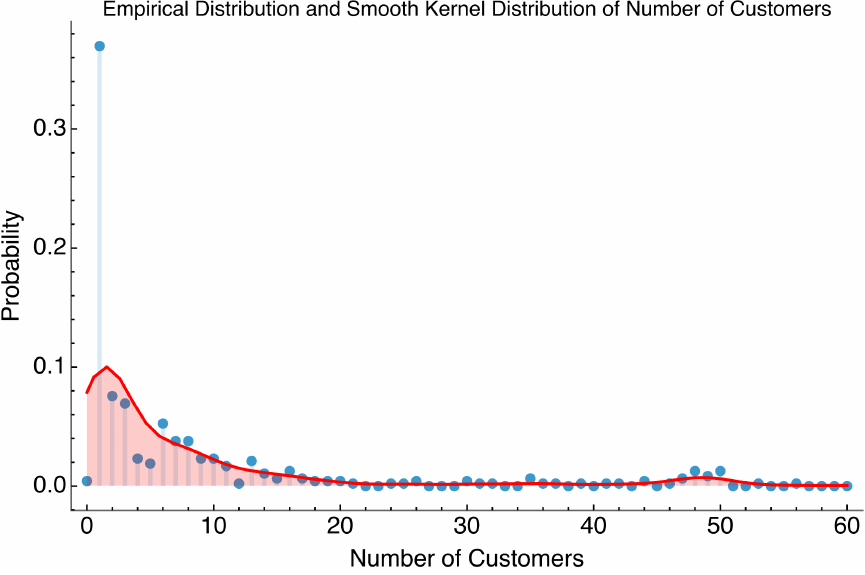}
    \caption{Empirical (blue) and estimated (red) probability density plot of the number of customers affected in overhead outages}
    \label{fig:ohCustomersPDF}
\end{figure}

Similar to the duration of overhead outages, the number of customers affected in overhead outages also has a highly right-skewed probability distribution, as shown in Fig. \ref{fig:ohCustomersPDF}. The estimated probability density using a Gaussian kernel is shown in red in Fig. \ref{fig:ohCustomersPDF}.
It has a skewness coefficient of 8.8.
While the duration of outages is a positive continuous random variable, the number of customers affected is a non-negative discrete random variable.

The mean, median, and mode of the number of customers affected in overhead outages are 23, 3, and 1, respectively. Again, we see that the median is a more effective summary statistic for representing the typical number of customers affected by overhead outages.

\section{Conclusions and Discussion}

We used a historical rerun method to quantify the benefits that the utility would have achieved if it had undergrounded 80\% of its overhead distribution circuits, and demonstrated the method using two overhead circuits. This data-driven method leverages the wealth of historical outage data readily available within a utility's Outage Management System. 
This historical data contains a comprehensive record of past outages, including all relevant factors that have demonstrably influenced their occurrence. 
These factors include weather events, system reconfigurations, infrastructure aging, accidents, load fluctuations, policy changes, and past resource/infrastructure upgrades.

The proposed approach uses historical data for simulated underground circuits to analyze how the undergrounded circuit would have performed during the same historical period. It can also be used to validate the benefits of undergrounding for lines that have already been converted from overhead to underground. By grounding the analysis in already observed data, the historical approach significantly reduces the uncertainties inherent in outage analysis compared to methods that rely on uncertain modeling with detailed physical models and assumptions about the future. This data-driven approach yields a more robust and reliable quantification of the potential benefits of undergrounding the distribution circuit.

Leveraging existing historical outage data stored within utilities' Outage Management Systems also offers several other advantages. Firstly, this data is readily available, obviating the need for additional data collection efforts. Secondly, the proposed approach is computationally efficient, requiring minimal computational resources. This enables straightforward implementation on existing utility infrastructure, significantly reducing the implementation burden for utilities.

The method relies heavily on the accurate estimation of the frequency, duration, and impact (i.e., the number of customers affected) of underground outages. This can be improved with more historical outage data and detailed information about the distribution circuits.
Alternatively, models of distribution cable outages could be used to estimate the underground outages.

\printbibliography

\end{document}